# Hyperspectral Infrared Microscopy With Visible Light

Anna V. Paterova[1], Sivakumar M. Maniam[2,4], Hongzhi Yang[1],
Gianluca Grenci[2,3]*, and Leonid A. Krivitsky[1]**

*[1]Institute of Materials Research and Engineering (IMRE),
Agency for Science Technology and Research (A\*STAR), 138634 Singapore*

*[2]Mechanobiology Institute, National University of Singapore, 117411 Singapore*

*[3]Department of Biomedical Engineering, National University of Singapore, 117583
Singapore*

*[4]National Institute of Education, Nanyang Technological University, 637616 Singapore*

\* mbigg@nus.edu.sg  \*\* Leonid_Krivitskiy@imre.a-star.edu.sg

*Hyperspectral microscopy is an imaging technique that provides spectroscopic information with high spatial resolution[1,2]. When applied in the relevant wavelength region, such as in the infrared (IR), it can reveal a rich spectral fingerprint across different regions of a sample. Challenges associated with low efficiency and high cost of IR light sources and detector arrays have limited its broad adoption. We introduce a new approach to IR hyperspectral microscopy, where the IR spectral map of the sample is obtained with off-the-shelf components built for visible light. The technique is based on the nonlinear interference of correlated photons generated via parametric down-conversion[3,4]. We demonstrate chemical mapping of a patterned sample, in which different areas have distinctive IR spectroscopic fingerprints. The technique provides a wide field of view, fast readout, and negligible heat delivered to the sample, which makes it highly relevant to material and biological applications.*

The development of optical metrology techniques in infrared (IR) range is quite active due to the richness of material and molecular signatures that are observable in this spectral range. Multiple IR spectroscopic instruments are commercially available but are saddled with non-ideal light sources and low-efficiency detector arrays. This limits the application for hyperspectral IR microscopy, in particular, for dynamic studies of live cells[5,6].

Typical light sources used for IR measurements are robust, low cost thermal blackbody emitters. Though their emission band covers the detection range of IR photodetectors, these sources have limited spectral brightness. Applications requiring bright and well-collimated beams, such as wide-field hyperspectral IR microscopy, uses light produced by synchrotron light sources[7] or high power quantum cascade lasers (QCLs)[8]. However, access to synchrotron light is limited and costly, while QCL technology has limitations in terms of cost, reliability, and spectral range / tunability. There are also limitations for light detection with wide-field IR microscopy. Currently, arrays of IR point detectors, referred to as focal plane arrays (FPAs) are used. These FPAs face several technical limitations such as cryogenic operation, high noise floor, non-uniform response, and thermal sensitivity[9]. They are costly and subjected to stringent end-user controls.

We propose and demonstrate a measurement to circumvent the need for IR sources and detectors in hyperspectral IR microscopy by employing quantum optical phenomena. Our method is based on the concept of the nonlinear interference of correlated photons, also known as *induced coherence*[3,4]. Two photons are generated in a nonlinear crystal via spontaneous parametric down-conversion (SPDC) with one photon (signal) in the visible range, and the correlated photon (idler) in the IR range[10]. The crystal is put into an interferometer in which the interference pattern of the detected visible photons carries information about the IR photons, which are the ones interacting with the sample. Information about the sample properties in the IR range is inferred from the measurements of visible range photons using standard visible light components.

Nonlinear interferometers have been used for IR metrology. Imaging in a transmission configuration has been demonstrated[11], but was limited to a single wavelength and a relatively low spatial resolution. Spectrally tunable optical coherence tomography is possible, but uses time-consuming point-by-point (raster) scanning[12,13]. Multiple experiments have also been performed with IR spectroscopy without imaging[14,15,16,17]. A related imaging scheme based on photon up-conversion requires the use of an expensive optical parametric oscillator[18].

In this work, we combine IR spectroscopy and microscopy in a single measurement for wide-field hyperspectral IR microscopy. We built a tunable IR microscope with high spatial resolution and fast readout and performed experiments in the mid-IR range, which is relevant to biological imaging and material investigation. As a proof of concept, we perform spectroscopic IR imaging and retrieve the absorption map of a microfabricated sample. The sample contains, prepared through UV-exposure of a photosensitive material, a spatial pattern with different chemical composition.

In our setup, the SPDC crystal is put in an imaging Michelson interferometer, where down-converted and pump photons are split according to their frequencies, see Fig.1. To realize a wide-field imaging configuration, we exploit the natural divergence of the SPDC photons[19,20]. We insert a three-lens system in each arm of the interferometer, which performs the Fourier transformation from a $k$-space into an $x$-space[21] (see Supplementary materials). The signal and pump photons are reflected by the reference mirror, while the idler photons are reflected by the sample. For the reflected beams, the lenses perform an inverse transformation, so that the spatial modes of the down-converted photons overlap in the crystal[22].

The state-vectors of down-converted photons, generated in the forward and backward passes of the pump through the nonlinear crystal, interfere. The intensity of the signal photons, measured in the experiment, is given by[12,23,24]:

$$I_s \propto 1 + |\mu| |t_i(\rho_{k_i})| cos\left(\varphi_p - \varphi_s(\rho_{k_s}) - \varphi_i(\rho_{k_i})\right), \qquad (1)$$

where $\rho_{k_s}$, $\rho_{k_i}$ are polar coordinates of signal and idler wavevectors, respectively, $|t_i(\rho_{k_i})|$ is the amplitude transmission for idler photons through the interferometer, $|\mu|$ is the normalized first-order correlation function of SPDC photons[4], and $\varphi_{p,s,i}$ are the phases of the pump, signal

and idler photons, respectively. Assuming that the interferometer is balanced ($|\mu| = 1$), the visibility of the interference pattern, is given by:

$$V(\rho_{k_s}) = |t_i(\rho_{k_i})|. \tag{2}$$

Thus, the transmission (absorption) map of the sample in the IR range can be inferred from the visibility map of the interference pattern of the visible range photons.

In the experiment, we measure the visibility of the interference pattern by scanning the relative phase between the interferometer arms. We tune the frequencies of the photons by changing the temperature of the crystal and selecting regions with different poling periods within the same crystal[25]. By capturing interference patterns of signal photons at different frequencies, we calculate the absorption of idler photons and perform the IR spectroscopic mapping of the sample, see Materials and Methods.

The sample under study is formed by the layer of SU-8 polymer (epoxy-based negative photoresist) coated on top of a (100) silicon wafer, see Fig. 2a. The thickness of the SU-8 layer is 21 μm, see Materials and Methods. SU-8 is a photo-sensitive polymer and its spectral response in the mid-IR range depends on the polymerization condition. For this experiment, we expose the photoresist film to UV light (i-line of a Hg arc lamp) through an optical mask with a spatial pattern of rectangles (exposed areas), while the surrounding area is kept non-exposed. The image of the sample under the optical microscope is shown in Fig. 2b. The image does not show any variance in the optical properties of different areas of the sample since SU-8 is transparent in the visible range. Note that rectangular edges of the exposed area can be seen due to slight shrinkage of the resist upon reticulation. To avoid the Fabry-Perot effect inside the SU-8 layer, we place a 400 μm thick calcium fluoride (CaF$_2$) window on top of the SU-8.

We probe the sample at different wavelengths of idler photons in the range from 2.75 μm to 3.35 μm with a step of ~25 nm. The correlated wavelengths of detected signal photons are in the range from 663 nm to 632 nm. The spectral width of the signal and idler photons is 1.91±0.02 nm and 42.8±0.5 nm, respectively (see Supplementary Materials). Examples of the absorption map of the sample at four probing wavelengths are shown in Fig. 3:

- Idler photon at 2.87 μm, signal photon at 653.2 nm. The exposed rectangles show stronger absorption than the surrounding non-exposed area, see Fig. 3a.
- Idler photon at 3.18 μm, signal photon at 638.8 nm. There is no distinctive contrast between different areas of the sample, see Fig. 3b.
- Idler photon at 3.32 μm, signal photon at 633.5 nm. The image inverts, and the exposed rectangles show weaker absorption than the non-exposed area, see Fig. 3c.
- Idler photon at 3.34 μm, signal photon at 632.8 nm. The contrast between different areas of the sample vanishes, see Fig. 3d.

The regions, indicated by dashed yellow circles, are air bubbles formed between calcium fluoride window and SU-8 interface, which are used as spatial references.

We set ranges of interest within corresponding areas on the sample (dashed black and red rectangles in Fig. 3) and measure the dependence of the absorption on the wavelength of idler photons. The results are shown in Fig. 4. The dots correspond to our experimental data, and the dashed curves show the FTIR data obtained for reference samples[26]. The FTIR data is normalized and averaged over a bandwidth of idler photons. The two data sets are in good agreement, which validates our method.

The accuracy of the visibility measurements in our experiments is 0.3%, which translates to the accuracy in the measurements of the absorption within 4%. The spectral resolution in our method is defined by the bandwidth of idler photons and is equal to ~43 nm. The spatial resolution is defined by the parameters of the imaging system and is equal to 50 μm. The spectral and spatial resolution can be tailored to a specific application by carefully choosing the parameters of the crystal and the magnification of the imaging system (see Supplementary Materials).

In conclusion, we demonstrated wide-field hyperspectral IR microscopy using off-the-shelf components designed for visible light. We measured the absorption pattern of SU-8 photoresist after parts of it were exposed by UV light. We tuned the wavelength of idler photons across the fingerprint region of the sample while detecting a correlated photon using a visible light CMOS camera. The IR absorption peaks, determined by our method, are consistent with the photo-induced reaction in the sample and with conventional FTIR data.

Our system simplifies IR wide-field microscopy while adding a hyperspectral component to this useful technique. Though the work within is in a reflection configuration, the method can be implemented in a transmission configuration by using alternative designs of nonlinear interferometers[11,27]. The method can also be used for polarization sensitive measurements[28]. With this demonstration, a new hyperspectral IR microscopy will be unlocked for practical applications in bio-imaging and material analysis.

## Materials and Methods

### *Experimental setup*
We use a 532 nm continuous wave laser (Laser Quantum, Torus, 80 mW) as a pump, see Fig.1. The laser beam is reflected by a dichroic mirror $D_1$ (Semrock), and then sent into the periodically poled lithium niobate (PPLN) crystal (1 mm × 10 mm × 10 mm) (HC Photonics) with three different poling periods of 9.84 μm, 10.28 μm and 10.73 μm. The crystal is put in the oven, controlled by a temperature controller (temperature range from 40 ˚C to 200 ˚C with the stability of ±0.1 ˚C). The oven is mounted on a linear stage to change the poling periods. Using three different periods and temperature tuning, we generate signal photons in the range of $\lambda_s$ = 663-632 nm and idler photons in the range of $\lambda_i$=2.75-3.35 μm wavelengths. Two lenses LS before the PPLN crystal form a collimated laser beam with a diameter of 1 mm, which fits into the clear aperture of the single poled region in the crystal.

The SPDC photons generated in the crystal are separated into different channels (visible and IR) by a dichroic mirror $D_2$ (ISP Optics). In each arm of the interferometer, three-lenses (uncoated CaF$_2$ bi-convex) are used. Lenses $F_1$ and $F_2$ have the focal distance of $f_{1,2}$=100 mm, while the lens $F_3$ has a focal distance of $f_3$=15 mm. The corresponding spatial resolution is about 50 μm. Results obtained with lenses with $f_3$= 25 mm, 15 mm, and 4 mm, are presented in Supplementary Materials. The pump and signal beams are reflected by mirror $M_s$, while the sample reflects the idler beam. The signal photons are filtered a notch and a bandpass interference filters (Semrock). The interference pattern is detected by focusing ($f$=150 mm) the signal SPDC photons into the silicon CMOS camera (ThorLabs, DCC1240C).

The sample is mounted onto a motorized *xyz* translation stage (ThorLabs), which allows one to image different areas of the sample, and balance the arms of the interferometer. The sample is attached to a piezo stage (Piezosystem Jena), which is used for fine scanning of the relative phase in the interferometer.

First, we balance the interferometer arms, and then we introduce fine phase shifts in the interferometer. At each position of the piezo stage, we capture the phase image of the sample. From the phase images, we infer the interference visibility, which is then used to calculate the absorption of the sample (see Supplementary Materials). The exposure time for each image is about 500 ms. We make 140 steps to cover the full period of the interference fringe. The overall acquisition time for each image at a given wavelength is about 70 seconds.

*Sample preparation*
The sample is prepared by UV-lithography. A 50 mm diameter standard (100) silicon wafer is used as a substrate, where we spin-coat a SU-8 3025 (Microchem, USA) photoresist with 21 μm thickness. After soft-baking (1 min at 65 °C followed by 10 min at 95 °C on hot plates), a periodic pattern of exposed rectangles (300 μm × 80 μm, distributed in a rectangular array with 130 μm and 160 μm gaps in two directions, respectively) is generated by UV-exposure (i-line of a Hg arc lamp, ~200 mJ/cm$^2$ power density at 365 nm) using a mask aligner (MJB4, Suss Micro Tec) and a soda-lime optical mask. Then, we perform post-exposure baking (5 min at 65 °C followed by 5 min at 95 °C). A change in the chemical composition of the exposed polymer occurs during the post-exposure baking. The 400 μm thick CaF$_2$ cylindrical window is placed on top of the photoresist during the post-bake.

*The spectroscopic response of the sample*

In the range of probing IR wavelengths, the absorption spectrum of SU-8 is defined by the absorption of –OH molecules and –C-O-C– epoxy rings. –OH groups present a distinctive broad peak of absorption at around 2.84 μm, while the epoxy ring has an absorption peak at 3.32 μm[26]. The absorption peaks assigned to –OH molecules and –C-O-C– epoxy rings follow the same behavior and present the characteristic switch of the relative absorption of the exposed and non-exposed SU-8 at 2.87 μm and 3.32 μm, see Fig.4. The observed dependence is correlated with the underlying chemical reaction induced by UV exposure. UV photons in the range around the i-line of the Hg arc lamp (340-350 nm) are absorbed by the photo-acid contained in the formulation of SU-8 photoresist (triarylsulfonium hexafluoroantimonate salts). This absorption activates the photo-acid through the formation of radical species. During the post-exposure bake, the radicals lead to the opening of the epoxy rings of the SU-8 monomers and the reaction propagates following the scheme of a classic acid-catalyzed cationic reaction that ends with the resin reticulation. This chemical process results in an increase of –OH molecules and a decrease of epoxy –C-O-C– rings in the exposed areas. Thus, the chemical modification induced by UV-exposure and post-baking is well reflected in our images shown in Fig. 3.

*Measurement of the absorption*

The amplitude transmission of the idler photons through our sample is given by $|t_i(\rho_{k_i})| = |\tau_i(\rho_{k_i})|^2 |r_i(\rho_{k_i})|$, where $|\tau_i(\rho_{k_i})|$ is the amplitude transmission coefficient of the sample for idler photons (single pass), and $|r_i(\rho_{k_i})|$ is the reflection coefficient of the back surface of the sample. The amplitude transmission coefficient of the sample $|\tau_i(\rho_{k_i})|$ is defined by (1) reflection from the air-CaF$_2$ surface $\sqrt{1 - |r_i^{SU8}|^2}$, (2) reflection from the CaF$_2$-SU-8 interface $\sqrt{1 - |r_i^{CaF2-SU8}|^2}$, and (3) transmission of the idler photons through the SU-8 $|\tau_i^{SU8}|$:

$$|\tau_i(\rho_{k_i})| = \sqrt{1 - |r_i^{CaF2}|^2}\sqrt{1 - |r_i^{CaF2-SU8}|^2}|\tau_i^{SU8}|. \qquad (3)$$

The reflection from the back surface $|r_i(\rho_{k_i})|$ is determined by the reflection of idler photons at the silicon-SU-8 interface $|r_i^{Si-SU8}|$:

$$|r_i(\rho_{k_i})| = |r_i^{Si-SU8}|. \qquad (4)$$

For calibration of the method, we use a gold-coated mirror as a reference. In this case, the visibility is defined by the reflectivity of the gold mirror $r_i^{Au}$:

$$V^{Au} = |r_i^{Au}|. \qquad (5)$$

Our reference measurements with a gold reflecting mirror yield the maximum visibility values of 30±0.3%, which is due to losses introduced by uncoated lenses in the three-lens system. The measured visibility maps with the sample are shown in Supplementary materials.

Taking into account equation (2), the ratio of the visibility of the interference pattern measured with the sample under study $V^{sample}$ in and the reference $V^{Au}$ is given by:

$$\frac{V^{sample}}{V^{Au}} = \frac{|\tau_i(\rho_{k_i})|^2 |r_i(\rho_{k_i})|}{|r_i^{Au}|} = (1 - R_i^{CaF2})(1 - R_i^{CaF2-SU8})T_i^{SU8}\frac{|r_i^{Si-SU8}|}{|r_i^{Au}|}, \qquad (6)$$

where $R \equiv |r_i|^2$ and $T \equiv |\tau_i|^2$ are the intensity reflection and transmission, respectively.

From equation (6) the transmission through the SU-8 photoresist is determined as:

$$T_i^{SU8} = \frac{V^{sample}}{V^{Au}}\frac{|r_i^{Au}|}{|r_i^{Si-SU8}|}\frac{1}{(1 - R_i^{CaF2})(1 - R_i^{CaF2-SU8})}. \qquad (7)$$

Using the Fresnel equations[29] and values of the refractive indices of the sample's components at the wavelength of 3 μm[30], we set the reflection from the air-CaF$_2$ surface at $R_i^{CaF2} = |r_i^{CaF2}|^2 \simeq 0.03$, the reflection from the SU-8 surface at $R_i^{CaF2-SU8} = |r_i^{CaF2-SU8}|^2 \simeq 0.003$, the reflection from the silicon-SU-8 interface at $|r_i^{Si-SU8}| \simeq 0.371$, and the reflectivity of the gold mirror at $|r_i^{Au}| = 0.99$. Then the absorption of the SU-8 is given by:

$$A_i^{SU8} = 1 - T_i^{SU8} = 1 - 2.76 \frac{V^{sample}}{V^{Au}}. \tag{8}$$

Here we neglect the wavelength dependence of the refractive index, its small difference between exposed and non-exposed areas of SU-8, and losses in $CaF_2$ window.

## Acknowledgments

We acknowledge the support of the Quantum Technology for Engineering (QTE) program of A*STAR and of Mechanobiology Institute (MBI-NUS). We thank Diego Pitta de Araujo from MBI Science Communications Unit for help in preparing some of the schematics and Reuben Bakker for insightful comments. We thank Kwek Leong Chuan from Centre for Quantum Technologies for helpful theoretical discussions.

## Figures and captions

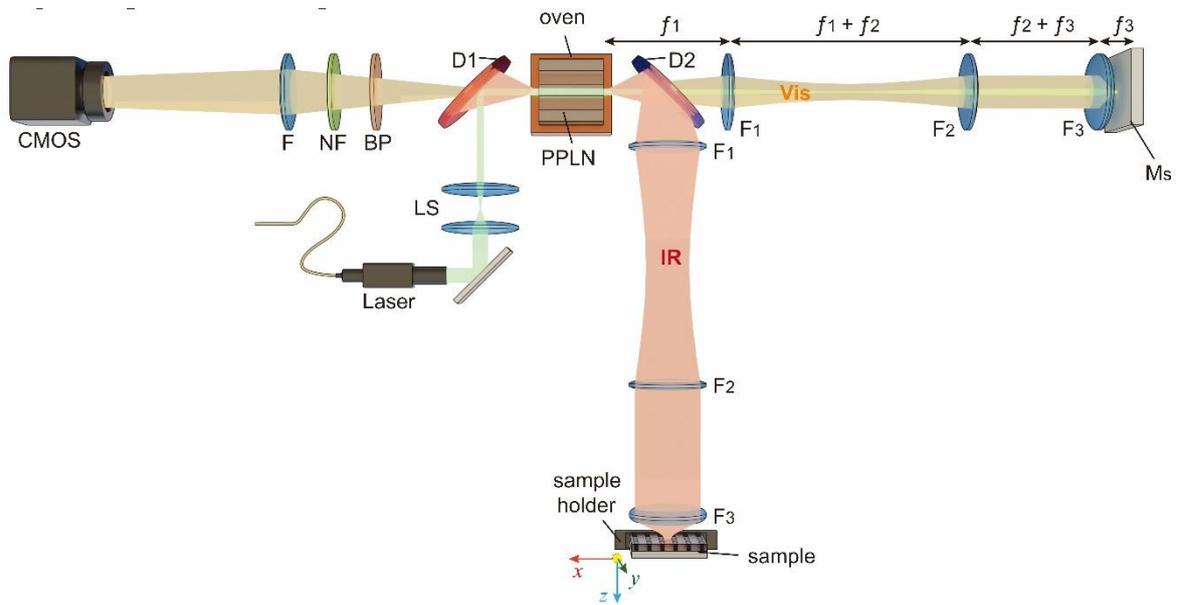

**Fig. 1: The experimental setup.** A continuous-wave laser is used as a pump for the PPLN crystal. The laser is injected in the setup by a dichroic beamsplitter D1. The signal (visible) and idler (IR) photons are generated via SPDC and separated into different channels of the Michelson-type interferometer by the dichroic mirror D2. The frequencies of the photons are tuned by changing the temperature of the crystal and revolving regions with different poling periods within the same crystal. The identical confocal three-lens systems $F_1$, $F_2$, $F_3$ in each arm of the interferometer project the $k$-spectrum of down-converted photons onto the reference mirror and the sample. The pump is reflected back to the crystal and generates other SPDC photons. The state-vectors of SPDC photons, generated in the forward and backward passes of the pump through the crystal, interfere. The interference pattern of signal photons is measured using a standard CMOS camera preceded by notch and bandpass filters (NF, F). The measured interference pattern of signal photons depends on the properties of the sample probed by idler photons.

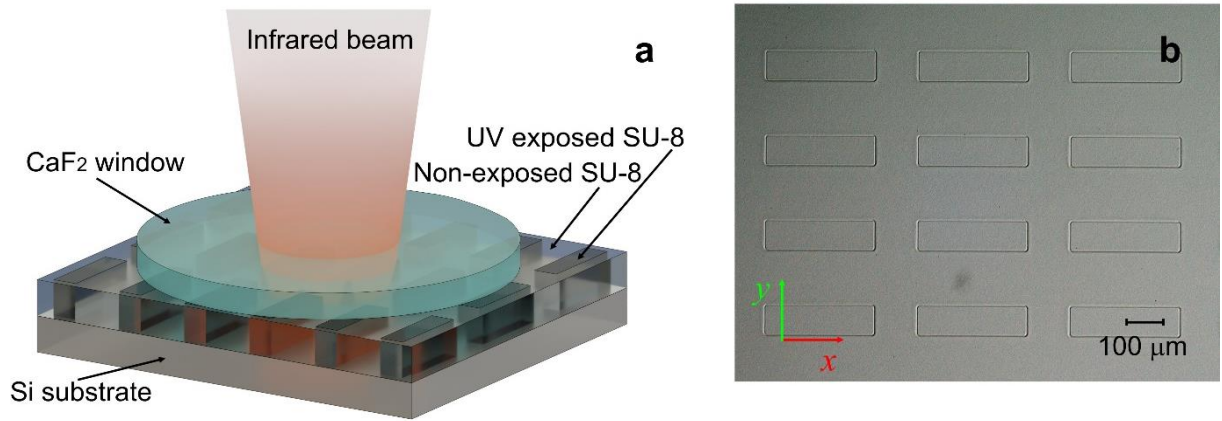

**Fig. 2: The sample. a** The schematic of the sample. A photo-sensitive polymer(SU-8) is coated on the silicon substrate. The CaF$_2$ window is placed on top of the SU-8 polymer to avoid the Fabry-Perot effect. UV exposed areas of the sample form rectangles (dark grey), which are surrounded by non-exposed SU-8. **b** The optical microscope images of the SU-8 layer through the CaF$_2$ window. Rectangles have 300 µm × 80 µm size and distributed in an array with 130 µm and 160 µm gaps in $x$ and $y$ directions, respectively.

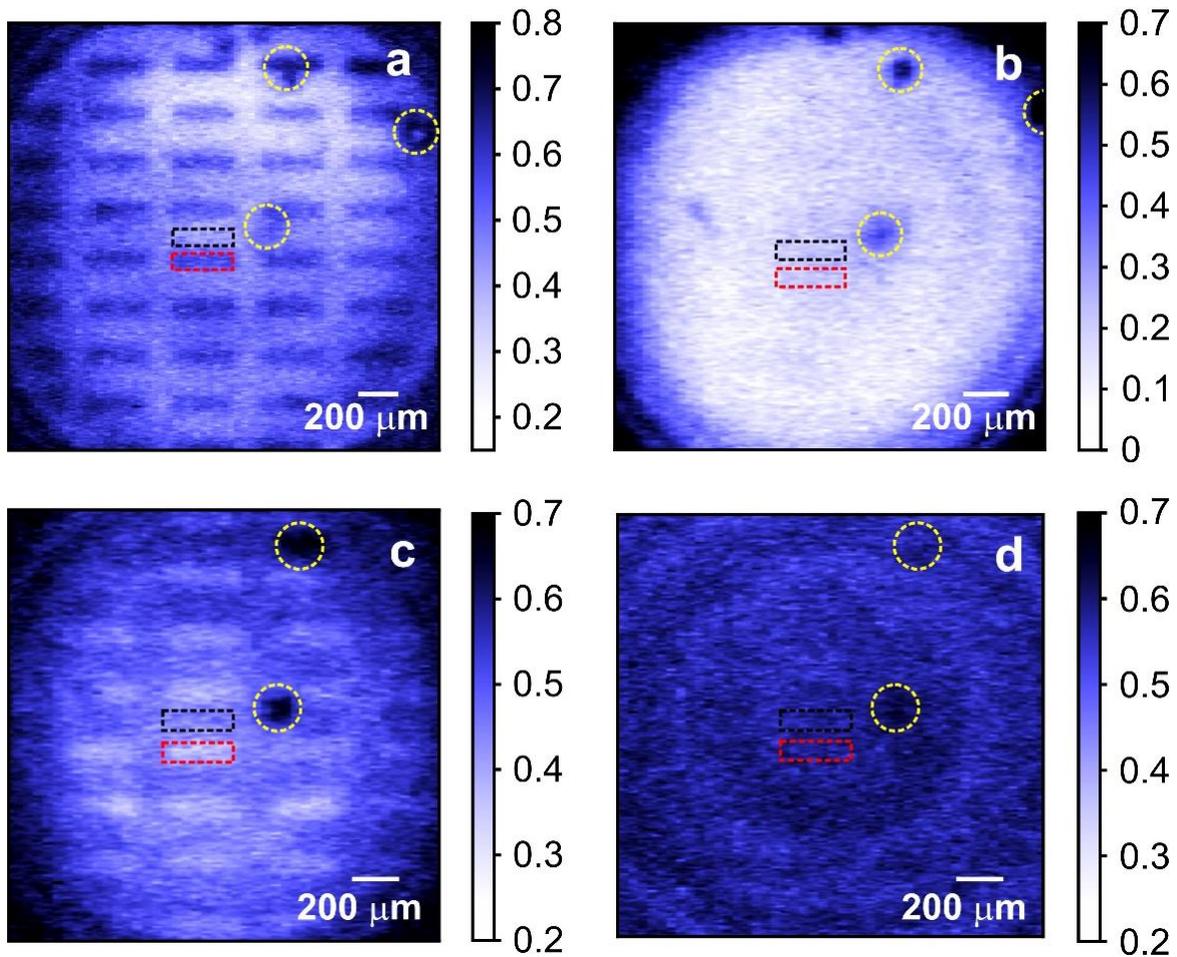

**Fig. 3: Hyperspectral IR microscopy**. Absorption map of the sample at probe (detected) wavelength of **a** 2.87 µm (653.2 nm), **b** 3.18 µm (638.8 nm), **c** 3.32 µm (633.5 nm), and **d** 3.34 µm (632.8 nm). The contrast between the exposed (rectangles) and non-exposed areas (surroundings) is changing depending on the wavelength of the probing IR photons, due to the difference in the chemical property triggered by UV illumination. The red (black) dashed rectangles show the exposed (non-exposed) areas of SU-8 used for the absorption measurements, shown in Fig.4. The dashed yellow circles highlight the air bubbles between $CaF_2$ and SU-8 interface, which were used as spatial references.

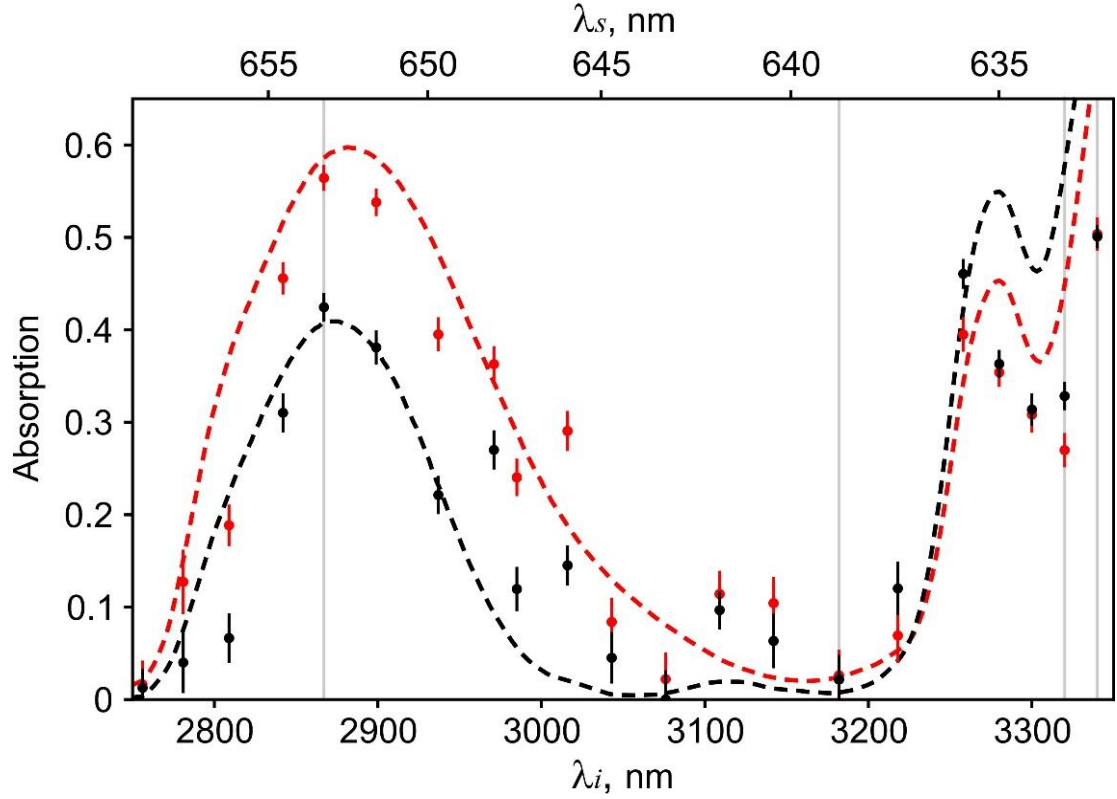

**Fig. 4 Measured IR absorption spectra.** Absorption spectra of SU-8 sample exposed (red curve) and non-exposed (black curve) by UV light. Dots show the experimental results; dashed curves show the FTIR data. Vertical gray lines show the wavelengths which correspond to images shown in Fig. 3. The bottom abscissa axis corresponds to the wavelengths of the idler probe photons, while the top abscissa axis shows the wavelengths of detected signal photons.

# Supplementary Materials

## *1. The optical configuration of the wide-field microscope*

The working principle of the imaging system is shown in Fig. S1. Each mode of the SPDC with a given $k$-vector is generated with a Gaussian intensity distribution[S1]. The spatial resolution of the system, is defined as follows:

$$\Delta\rho_k = \frac{\lambda_i}{2\,NA} = f_1\frac{\lambda_i}{2nr},\qquad\text{(S1)}$$

where $\lambda_i$ is the wavelength of the idler photons, $NA$ is the numerical aperture, $f_1$ is the focal length of the lens, $r$ is the radius of the beam, $n$ is the refractive index of the medium where the beam propagates.

The corresponding field of view (FOV) of the imaging system is given by:

$$FOV = 2f_1\theta_{max},\qquad\text{(S2)}$$

where $\theta_{max}$ is the maximum emission angle of the SPDC (within a given spectral range), typically a few degrees, where one can still observe interference pattern.

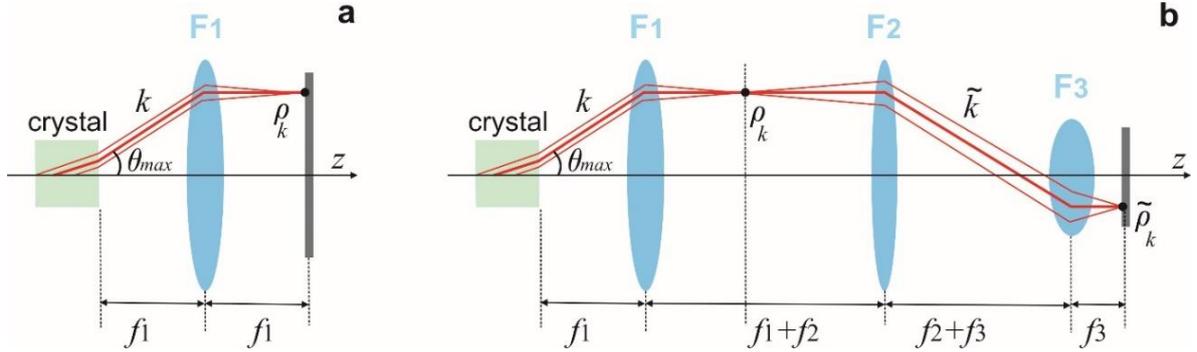

**Fig. S1:** The imaging system in one of the arms of the interferometer consisting of **a** one and **b** three lenses. In **a,** the crystal and the sample are placed at a focal distance from the lens $F_1$. In **b,** lenses $F_1$, $F_2$ and $F_3$ with their respective focal length of $f_1$, $f_2$, and $f_3$ are positioned confocally. The parameters of the three-lens system define the spatial resolution of the method. The beams reflected from the sample (and reference mirror) take reciprocal paths as the original beams emitted from the crystal.

Equation (S1) suggests that a higher resolution can be achieved by decreasing $f_1$. However, choosing the lens with small $f_1$ is not practical, because it does not allow placing a dichroic beam splitter after the crystal. Hence, we insert two additional confocal lenses to achieve the desired resolution, see Fig. S1b. The first lens $F_1$ is placed at a focal distance $f_1$ from the crystal; the second lens $F_2$ with the focal distance $f_2$ is placed at a distance $f_1+f_2$ from the first lens, and the third lens $F_3$ with the focal distance $f_3$ is placed at a distance $f_2+f_3$ from the second lens. The sample is placed at the focal distance of the third lens $f_3$. The three-lens system in each arm of the interferometer projects the $k$-spectrum of signal and idler photons on the mirror and the sample, respectively.

The resolution and the field of view after the three-lens system are given by:

$$\begin{cases} \Delta \widetilde{\rho}_k = \frac{f_1 f_3}{f_2} \frac{\lambda_i}{2nr} \\ \widetilde{FOV} \approx 2 \frac{f_1 f_3}{f_2} \theta_{max} \end{cases}, \tag{S3}$$

According to equation (S3), smaller $f_1$ and $f_3$, and larger $f_2$ are required to achieve the higher resolution.

## 2. *Experimental study of the spatial resolution*

We study the spatial resolution of our technique by imaging a reflective chromium-coated resolution test target (USAF 1951, ThorLabs). We scan the relative phase between two arms of the interferometer and acquire spatially resolved interference patterns. Each pattern is acquired for 500 ms. The wavelengths of the idler and signal photons are set to 3 μm and 647 nm, respectively.

Fig. S2a,b shows the interference pattern acquired at the relative phase $\varphi_i = 0$, and $\varphi_i = \pi$, respectively. The contrast between the two patterns is proportional to the reflectivity of the sample. Uncoated regions of the sample (glass substrate) show lower contrast than the background (chromium coated). The difference can be seen in Fig. S2c,d, which shows interference fringes form the chromium coated (red pixel in Fig. S2a,b) and glass surfaces (gray pixel in Fig. S2a,b), respectively.

Next, we can plot the contrast (visibility) maps across the sample, see Fig. S3a. By comparing the obtained visibility map with the reference, acquired using a gold mirror as a sample, see Fig. S3b. Then we infer the reflectivity of the sample, see Fig. S3c.

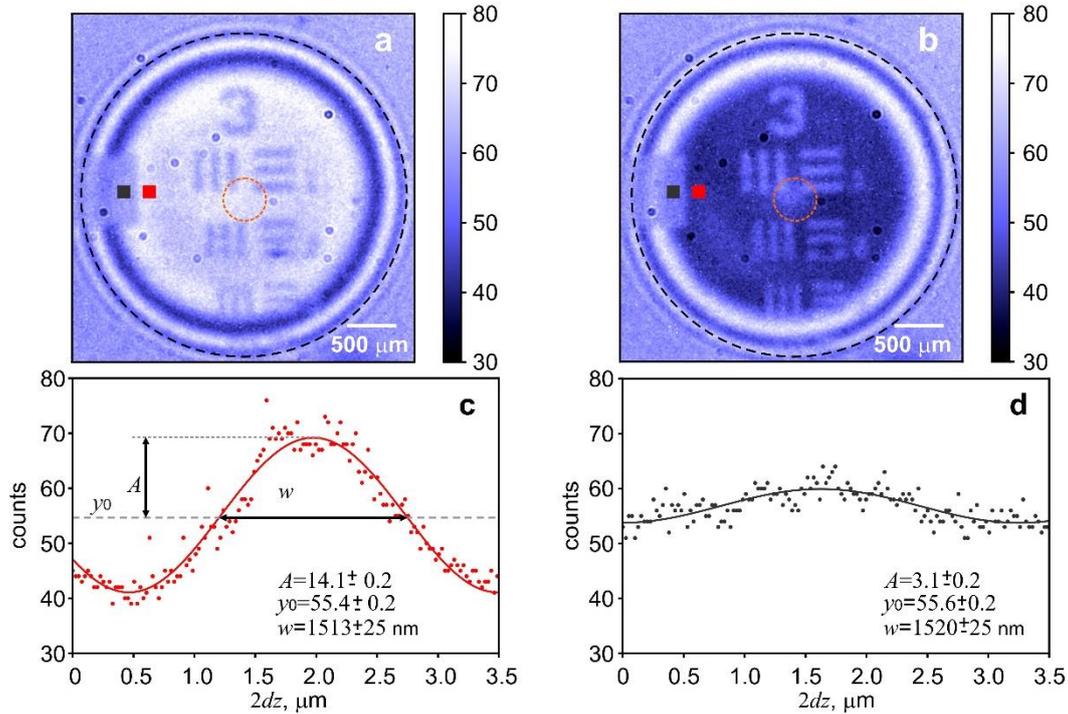

**Fig. S2:** Single interference patterns produced by the resolution test target at the relative phase **a** $\varphi_i = 0$ and **b** $\varphi_i = \pi$. The patterns are obtained using the lens $F_3$ with focal length of $f_3$=15 mm. **c** and **d** show interference fringes measured as a function of the path length difference for the reflective chromium region (red pixel in **a,b**) and the glass substrate (gray pixel in **a,b**), respectively. The wavelength of the probe photon is 3 μm and the detected photon is 647 nm.

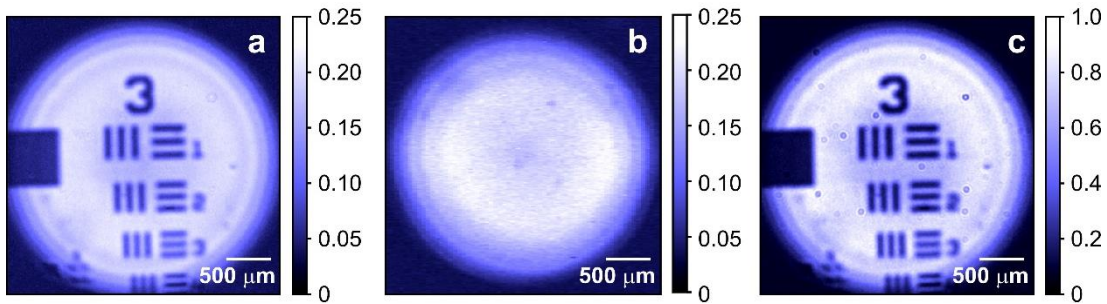

**Fig. S3: a** The visibility map of the interference pattern obtained using the resolution test target obtained from 140 interference patterns, see Fig. S2. **b** The reference visibility map acquired with the gold mirror **c** The inferred reflectivity of the resolution test target. The wavelength of the probe photon is 3 μm and the detected photon is 647 nm.

Next, we image the resolution test target with lenses $F_3$ of different focal lengths. Figure S4 shows the results obtained with $f_3$=25 mm. Here, four measurements are stitched together to obtain the full image of the resolution test target. Insets show fragments of the image obtained using lenses with $f_3$=15 mm and $f_3$=4 mm focal distance. For lenses with focal distances $f_3$=25 mm, $f_3$=15 mm and $f_3$=4 mm we obtain the following values of the spatial resolution: 78 μm, 50 μm and 15 μm, respectively, which are close to the theoretical estimates given by equation (S3).

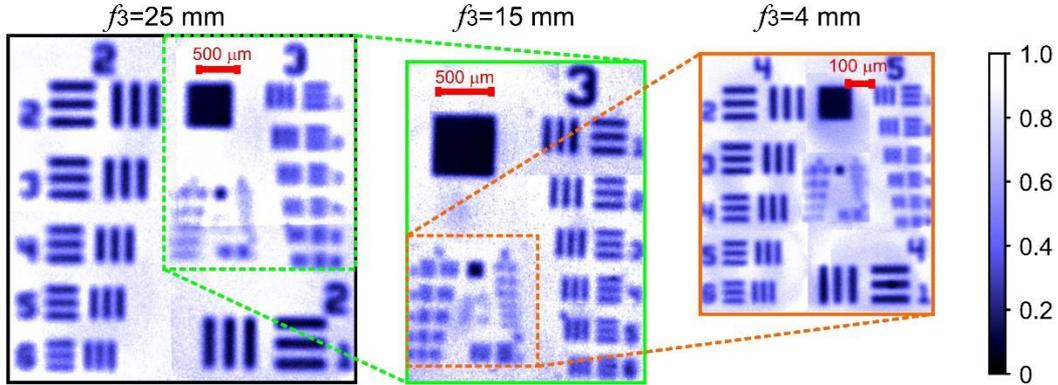

**Fig. S4:** The image of the resolution test target measured using lenses F$_3$ with different focal distances $f_3$. The image on the left corresponds to the $f_3$=25 mm; insets in the center and on the right show results for $f_3$=15 mm and $f_3$=4 mm, respectively. The scale shows the reflectivity of the sample.

### 3. *The spectrum of SPDC photons*

The measured spectrum of the signal photons is shown in Fig. S5. Orange triangles show the spectrum in the central region of the interference pattern which correspond to the orange dashed circle in Fig. S2a,b. The black dots show the spectrum integrated across the whole detection region, shown by the black dashed circle in Fig. S2a,b. Figure S5 illustrates that the central wavelength of the SPDC λ$_0$ varies depending on the scattering angles. The full width at the half maximum of the central spot is 1.91±0.02 nm with the corresponding linewidth of the correlated IR photon calculated to be 42.8±0.4 nm.

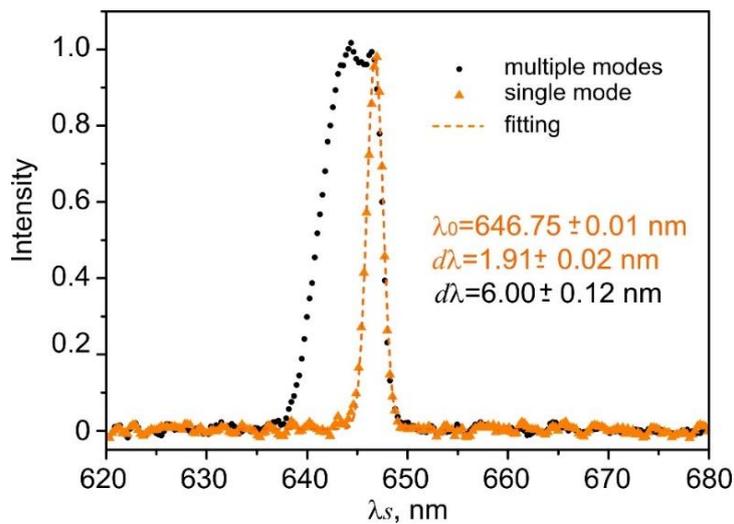

**Fig. S5:** The normalized spectrum of the signal photons obtained in different areas of the detected interference pattern. Orange triangles show the spectrum obtained at the center of the image, and black

dots show the spectrum obtained across the full field of view. Both regions are shown in Fig. S2a,b by orange and black circles, respectively.

4. *Raw visibility data*

At the experiment we obtain the visibility of the interference pattern first. The raw visibility data at the three probing wavelengths are shown in Fig. S6. Then, based on the visibility data we retrieve the absorption map of the sample.

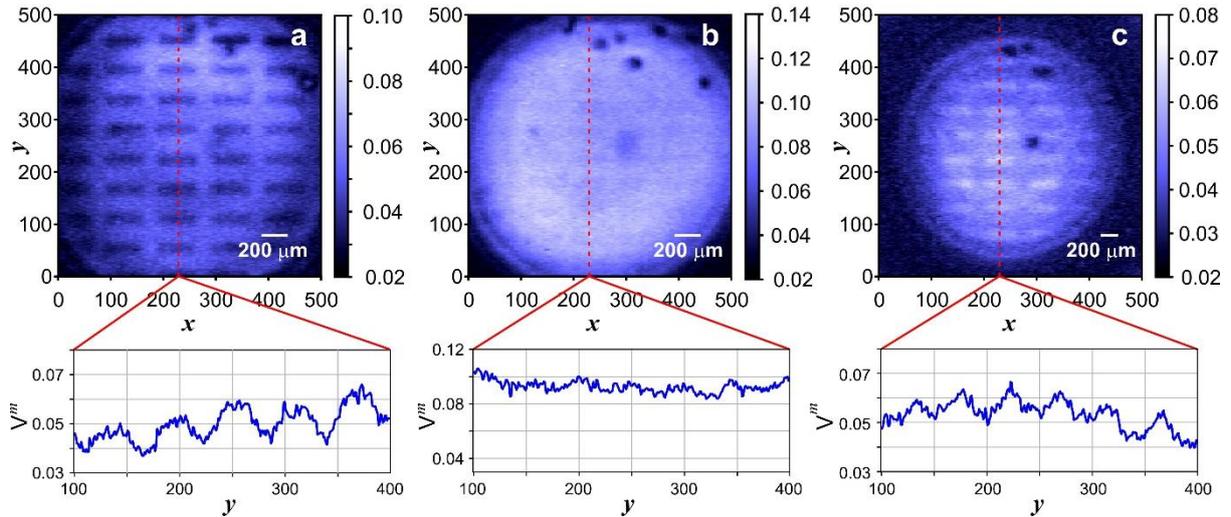

**Fig. S6:** Visibility map of the interference at **a** 2.87 µm, **b** 3.18 µm and **c** 3.32 µm probe wavelengths. The red dashed lines correspond to the cross-sections shown at the bottom of each picture. The horizontal axis in inserts are number of pixels of the CMOS camera.